\documentstyle[aps,prl,psfig]{revtex}
\begin{document}

\title{A simple one-dimensional model of heat conduction which obeys Fourier's law}
\author{P.L. Garrido$^1$, P.I. Hurtado$^1$ and B. Nadrowski$^2$}
\address{
$^1$  Departamento de E.M y F{\'\i}sica de la Materia\\ Facultad de Ciencias,
Universidad de Granada, 18071 Granada, Spain.\\
$^2$Institut Curie, UMR 168, 11, rue Pierre et Marie Curie, Paris Cedex 05, France. 
}
\date{\today}
\maketitle
\begin{abstract}
We present the computer simulation results of a chain of hard point particles
with alternating masses interacting on its extremes with two thermal baths at different temperatures.
We found that the system obeys Fourier's law at the thermodynamic
limit. This result is against 
the actual belief that one dimensional systems with  momentum conservative
dynamics and nonzero pressure have infinite thermal conductivity. It seems that  
thermal resistivity occurs in our system due to a cooperative behavior in which light particles
tend to absorb much more energy than the heavier ones.
\end{abstract}
J.B.J. Fourier developed his theory of heat conduction at the beginning of 19th century.
It states (in actual notation) that the temperature profile of an isolated system 
will evolve following the conservation equation
\begin{equation}
c_v(T){{\partial}\over{\partial t}}T(\underline{r},t)=\underline{\nabla}\cdot\left[
\kappa\underline{\nabla}T\right]
\label{Fourier}
\end{equation}
where $T(\underline{r},t)$ is the temperature measured by a probe at position $\underline{r}$ at
time $t$, $c_v(T)$ is the specific heat per unit volume and $\kappa$ is the thermal conductivity.
Fourier's law may be applied, in particular, to a system in contact with two heat 
reservoirs at different temperatures placed at $x=0$ and $x=L$. In this case,
the stationary state has the property of
\begin{equation}
J=-\kappa(T){{dT}\over{dx}}=const.
\end{equation}
where $J$ is the stationary heat flux through the system. Notice that one should assume that
there is not mass transport and/or other mechanisms different than heat conduction.
This law has been extensively tested in experiments in fluids and crystals. However,
we do not understand yet many of its fundamental aspects
(see for instance the review in \cite{Joel}).
In particular, its derivation from a microscopic Hamiltonian dynamics and
the existence of a local equilibrium that gives meaning to the local temperatures are  
open problems.

Actually, heat transport in one dimensional systems is  an interesting problem in the
context of both
nonlinear dynamics and statistical physics.
Long ago, Peierls proposed a successful perturbative theory, based on a phonon scattering 
mechanism, in order to explain the thermal conductivity in solids\cite{Peierls}.
In particular, Peierls theory  predicts that we do not expect a finite thermal conductivity
in one-dimensional monoatomic lattices with interactions between nearest neighbors.
Accordingly, the temperature
profile of a chain of $N$ harmonic oscillators is flat and its thermal conductivity goes like
$\kappa(N)\simeq N$ for large enough $N$. More generally, any integrable 
Hamiltonian system is expected to have
such divergent conductivity because its associated normal modes behave like a gas
of non-interacting particles carrying energy from the hot source to the cold one with
no loss. 
On the other hand, there are non-integrable systems, to which the Peierls theory does not
apply directly, whose behavior is known to agree with its prediction.
For example, the thermal conductivity of the Fermi-Pasta-Ulam-$\beta$ model goes
like $\kappa(N)\simeq N^{\alpha}$, with $\alpha=0.43$. The value of $\alpha$ can be 
predicted by using a mode coupling approximation to the corresponding interacting
gas of phonons \cite{Livi}. 
An exception to these results are systems with translation invariant Hamiltonian that
have zero pressure. For instance,  a one dimensional chain of rotators show
normal heat conduction \cite{Giardina}. 

In general, as an extension of the prediction from Peierls argument, it is presently
believed that
one shouldn't expect in general a finite thermal conductivity in
one-dimensional systems with momentum conserving interactions and nonzero
pressure\cite{Prosen}.
The goal of this letter is to show a counterexample to the above belief. 
We introduce a system that, although its particle interaction conserves
momentum and the pressure is non zero, the energy behavior has a diffusive character 
and Fourier's law holds.
Therefore, we think that in one dimensional systems with nonzero pressure, 
the conservation of momentum 
does not seem to be  a key 
factor to find
anomalous heat transport.

\begin{figure}
\centerline{
\psfig{file=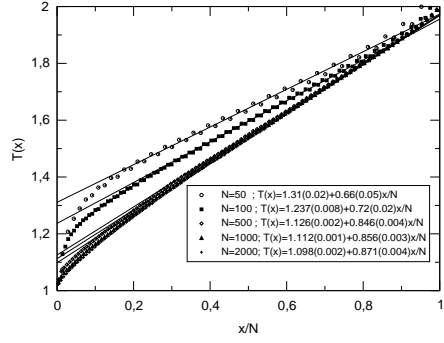,width=9cm,angle=-90}}
\caption{Temperature profile at the stationary state for $N$ particles. Lines are the
best fits of the data in the interval $x/N\in[0.4, 0.6]$. The corresponding equations are 
shown in the box.
Errors in the coefficients are in brackets.}
\label{figura1}
\end{figure}

Let us introduce our model. In a line of length $L$, there are 
$N$ point particles of different masses interacting
exclusively via elastic collisions. In order to minimize the finite size effects, the particles
have only two different masses and they alternate along the line, {\it i.e.} 
$m_{2l-1}=1$ and $m_{2l}=(1+\sqrt{5})/2$ with $l=1,...,N/2$. We have chosen 
the masses of the even particles to be the most irrational number in order to minimize 
possible periodicities, resonances or non-ergodic behaviors. 
At the extremes of the line there are thermal reservoirs at fixed temperatures 
$T_1=1$  and $T_2=2$ at $x=0$ and $x=L$, respectively. We simulate the reservoirs 
by using the following process: each time particle $1$ ($L$) hits the boundary 
at $x=0$ ($x=L$) with velocity $v$, the particle is reflected with the velocity modulus
\begin{equation}
v'=\left[ {{-2}\over{m_{1(L)}\beta_{1(2)}}}\ln\left(1-e^{-{{\beta_{1(2)}}\over{2}}m_{1(L)}v^2}
\right)\right]^{1/2}
\end{equation}
where $\beta_{1(2)}=1/T_{1(2)}$.
This reversible and deterministic map is due to H. van Beijeren (private 
communication). In order to check the influence of the type of reservoir into the system 
properties we have also used more conventional stochastic boundary conditions, but 
only different finite size effects and no other relevant behavior has been observed.
For $T_1\neq T_2$ there is 
a flow of energy from the high temperature reservoir to the low one, and the system then 
evolves to a non-equilibrium stationary state.
A version of this model in which the masses are randomly placed was already studied
by us \cite{garrido}.  In this work, the system thermodynamic 
limit behavior was not considered but
the local equilibrium property was demonstrated. 
Our goal in this letter is to check if the system has a finite thermal conductivity in the 
thermodynamic limit $N,L\rightarrow\infty$ with $N/L=1$.
With this aim we performed 
a detailed numerical  analysis along the following lines:

{\it(1) The existence of a nontrivial thermal profile:}
We define the local temperature by measuring the mean kinetic energy of each 
particle and its mean position at the stationary state. 
We computed the profiles for $N=50$, $100$, $500$, 
$1000$, $2000$, with fixed $N/L=1$, $T_1=1$ and $T_2=2$. Fig.\ref{figura1}  
shows the local temperature as a function of $x/N$
(by seeking clearer figures, we have performed local averaging of the temperatures and 
positions to draw only $100$ 
points; no difference is found by drawing all the points). 
We see in Fig.~\ref{figura1}  that the temperatures follow linear profiles in the
interval $x/N\in[0.4, 0.6]$ with slopes 
depending on the system size. This slope  apparently tend to converge to unity but we 
find that the convergence is so slow that we cannot (numerically) exclude the 
possibility that the limiting temperature profile is nonlinear (which is not against
Fourier's law). In any case, a non-flat profile is clearly expected in the thermodynamic limit.

{\it (2) The averaged heat current:}
If Fourier law holds and the heat conductivity is 
finite, the mean heat current, $J=N^{-1}\sum_{i=1}^N m_iv_i^3/2$ should go to zero as 
$1/N$ whenever $T_{1,2}$ and $N/L$ are kept fix. The data does not give us
a conclusive answer.
In fact we fitted the experimental
points ($J$ corresponding up to seven different $N$'s) to behaviors like $J=aN^{-0.71}$,
$J=aN^{-1}(1+bN^{-1})$, $J=aN^{-1}(1+b\ln N)$ and $J=aN^{-1}(1+b/\ln N)$ all of
them with regression parameters of order $0.999$. This reflects that the corrections to
the leading order are dominant and that we are far from the asymptotic regime for the
observable heat current\cite{dhar}.
Therefore,  the
direct use of the Fourier Law $\kappa=J N/(T_2-T_1)$ does not clarifies (from the
numerical point of view) the existence of a finite heat conductivity 
in the thermodynamic limit.

\begin{figure}
\centerline{
\psfig{file=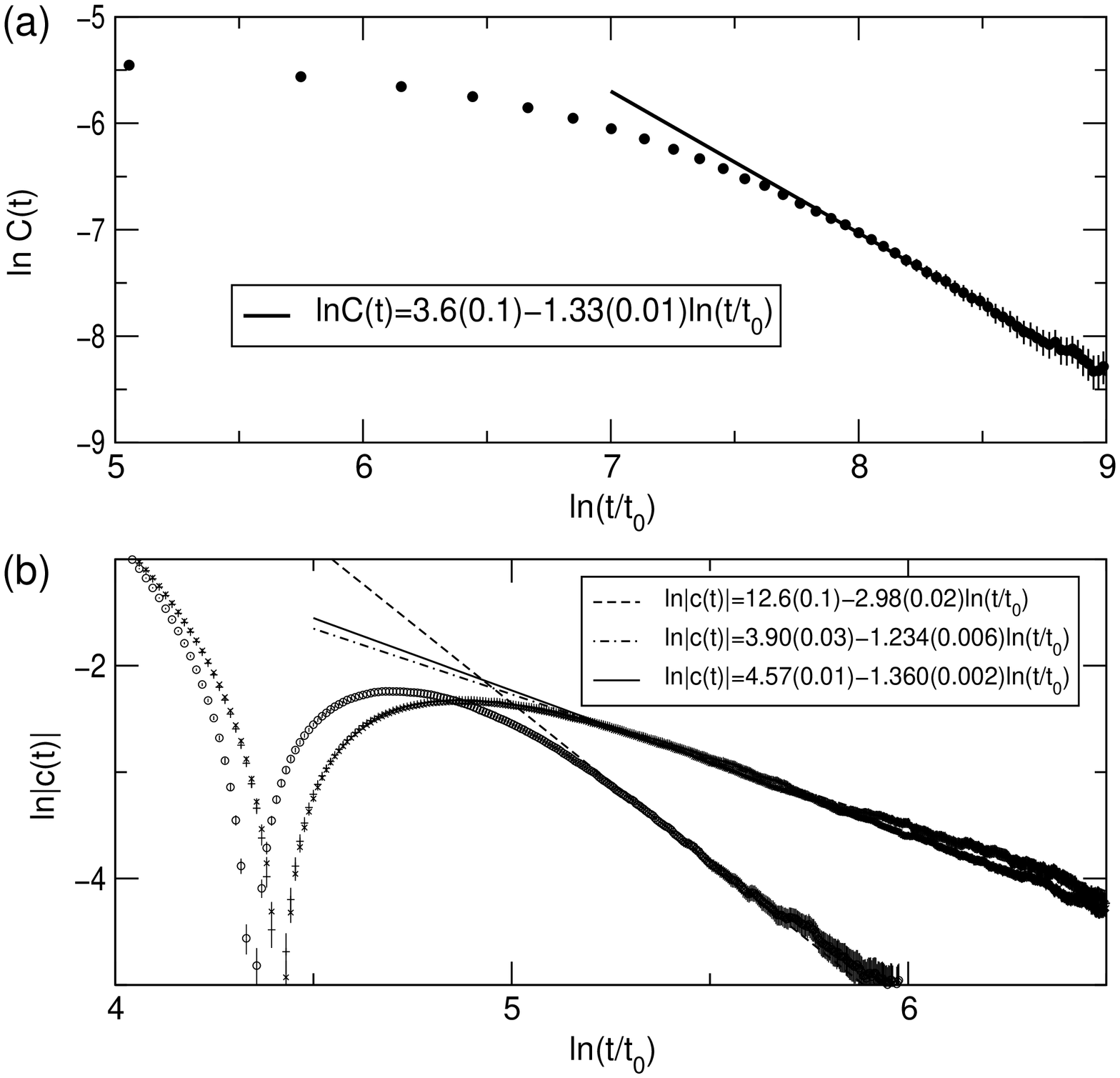,width=9cm,angle=-90}}
\caption{(a) Logarithm of the total heat current self correlation function versus
logarithm of t. The data corresponds to a system with $N=500$. Solid line is
the best fit for the asymptotic region. The number of independent averaged histories
is of order $10^7$.\\
(b) Logarithm of the absolute value of the local heat current self correlation
function versus logarithm of t (see text). ($+$) and ($\times$) symbols correspond 
to a system with $N=500$ and $N=1000$ respectively. ($\circ$) are the results for
a system of $N=500$ particles with equal masses.
Lines are the best fits to the asymptotic
regions. Their equations are shown in the box. Slope $2.98(0.02)$
correspond to the equal masses case. Slopes
$1.234(0.006)$ 
and $1.360(0.002)$ correspond to $N=500$ and $N=1000$ with different
masses, respectively. In all cases, the number of independent averaged histories  is
of order $10^9$.}
\label{figura2}
\end{figure}

{\it  (3) The current-current self correlation function:} The heat 
conductivity is connected to the current-current self correlation function evaluated 
at equilibrium via its time integral (Green-Kubo formula). The integral has some 
meaning whenever the correlation function decays
as $t^{-1-\Delta}$ with $\Delta>0$. 
We measured the current current correlation function with
periodic boundary conditions and total momentum equal to zero. We find that
the simulation has strong finite size effects. Nevertheless we can
see a clear tail of order $t^{-1.3}$ (see (a) in Fig.~\ref{figura2}). 
In order to confirm such result
we computed the local current-current correlation function that it has much
better averaging properties.
We show in Fig.~\ref{figura2} (b) the behavior of
the logarithm of 
$c(t)=\langle j_i(0)j_i(t)\rangle$ with $j_i(t)=m_i*v_i(t)^3/2$,
where the average is evaluated at the equilibrium state with $T_1=T_2=1.5$ and we
average over all the particles and different initial states.
In order to check that we are doing right we first computed $c(t)$ when the system has 
equal masses. In this case we know from Jepsen that the exact solution \cite{Jepsen} behaves
like   
$c(t)\simeq t^{-3}$ for $t$ large enough.
In Fig.~\ref{figura2} (b) we see how this behavior is obtained numerically and we also see
that for the different masses case $c(t)$ decays as  $t^{-1-\Delta}$ where $\Delta$ is again  
close to  $0.3$. 
This implies that we can define a finite thermal conductivity via 
Green-Kubo. We think that the decay of correlations is so slow
that it explains the strong finite size effects observed in the temperature 
profile and in the mean heat current. In fact we can argue that 
$J N/(T_2-T_1)=\kappa-A N^{-\Delta}$ which explains why
we do not see a clear behavior of $J$ with $N$ with system sizes of order $10^3$
(the corrections are of order unity  for those sizes).

\begin{figure}
\centerline{
\psfig{file=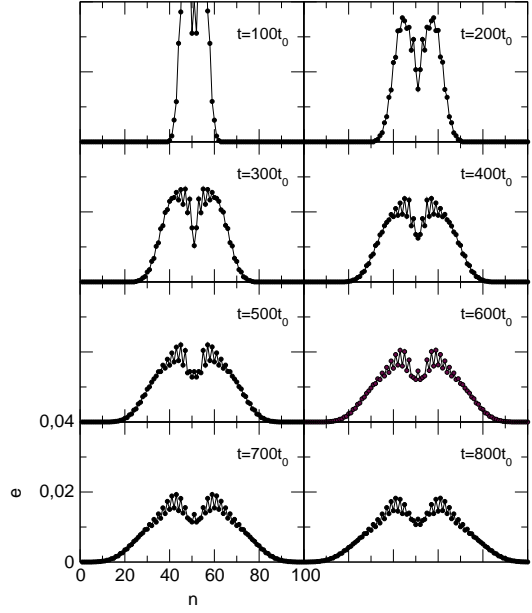,width=11cm,angle=-90}}
\caption{Evolution of the energy distribution for an initial condition in which all particles,
$N=100$,
are at rest except particle $51$ which has an averaged energy corresponding to temperature
$1.5$. The figure shows averages over $10^7$ independent realizations and
$t_0=0.032$.}
\label{figura3}
\end{figure}

{\it (4) The energy diffusion:} We also wanted to check if the dynamical version 
 (see Eq.\ref{Fourier}) of Fourier's law holds. With this aim,  we prepared the system with 
zero energy (all particles at rest) and positions $x(i)=i-1/2$, $i=1,\ldots, N$. 
Then, we give to the light particle $i=N/2+1$ a velocity
chosen from a Maxwellian distribution with temperature $T=1.5$. We monitored how the
energy flows through the system until any boundary particle moves. Finally, we average 
over many initial conditions. If the system
follows the Fourier's law we should see a diffusive type of behavior (if the
thermal conductivity is constant). Figure \ref{figura3} shows the energy distribution
for $N=100$ and different times measured in units $t_0=0.032$. 
Let us remark here that to apply Eq.\ref{Fourier} the  temperature should have
a smooth variation in the microscopic scale to guarantee that local equilibrium holds.
In Figure \ref{figura3} we see that, for times larger than $t=200t_0$, the average
variation in the local temperature is of order $0.001$. Therefore, we may assume that
we are in a regime where Eq.\ref{Fourier} holds.
Initially, the energy of the
light particle is transfered to the neighbors very fast and then the particle stays very cold, much 
colder than its neighbors. In fact, in this initial regime, the energy maxima are  moving outwards 
at constant velocity. This behavior ends at around $t=100t_0$. The system then
begins to slow down and, at $t\simeq 300t_0$, the structure of the energy distribution 
changes, and one
can then differentiate the behavior corresponding to light particles and heavy ones at least
around the maxima of the distribution. We measured the mean square displacement of
the energy distribution at each time: $s(t)=\sum_n (n-51)^2 e(n,t)$. We found that we can
fit $\ln s(t)=-6.39(0.04)+2.05(0.01)\ln t$ for $t/t_0\in(30,100)$, thus, a ballistic behavior 
that changes smoothly
until for $t>400t_0$ were we find a diffusive behavior $\ln s(t)=-1.00(0.01)+1.005(0.002)\ln t$. 
This last result confirms that our system follows even the dynamical aspects of Fourier's law.

\begin{figure}
\centerline{
\psfig{file=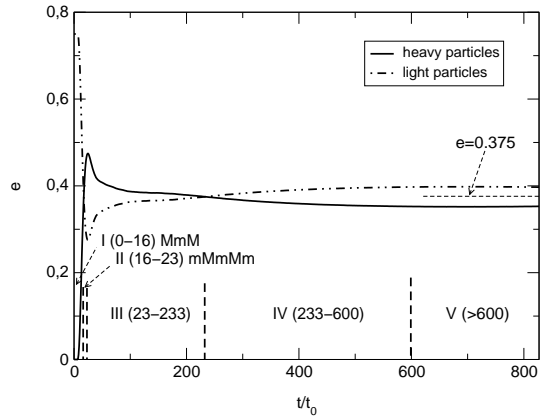,width=10cm,angle=-90}}
\caption{Evolution of the total energy stored in the heavy and light particles. 
The conditions are the same as in Fig.~\ref{figura3}. $MmM$ indicates that only the central light
particle and the nearest heavy ones are moving in region I. }
\label{figura4}
\end{figure}

As we noticed above, in Fig.~\ref{figura3} we see that the light  and heavy particles seem
to follow different energy distributions, at least for times longer than $t=300t_0$. 
In order to get some more insight about such behavior,
we computed the evolution of the total energy stored in the light (heavy) particles.
The result is shown in Fig.~\ref{figura4} where 
we can detect five different time regions:
{\it(I) $t/t_0\in(0,16)$}; only the light particle and the two heavy nearest neighbors 
have a
nonzero velocity. {\it (II) $t/t_0\in(16,23)$}; the five central particles (three light and two heavy ones)
are moving. The total energy stored in the light particles reaches a minimum.
{\it (III) $t/t_0\in(23,233)$}; the heavy particles begin  to release energy (on the  average) 
until, at $t=233t_0$, both types of particles have
the same amount of energy. {\it (IV) $t/t_0\in(233,600)$} Light particles keep
getting energy until we reach region. {\it (V) $t>600t_0$}; where the total energy stored in the
light  particles reaches a constant value  that exceeds to the one corresponding to the 
heavier ones.
Let us remark  that, in the asymptotic regime $t>600t_0$,
the energy distribution is still evolving and, therefore, this partition of energy between
both degrees of freedom is an asymptotic dynamical property of the system.

In order to discard any non-ergodic behavior of our
system we included reflecting boundary conditions at the extremes of the chain and we
did much longer simulations. We saw that the isolated system tends to the equilibrium
in which equipartition of energy between all degrees of freedom holds. 
That is, the total energy stored in the light particles is
equal to the one stored in the heavy ones. 
Moreover, we have checked that the system at any stationary state 
(equilibrium or non-equilibrium) does not 
present the property of non-equipartition of the energy.  
We think that this non-equipartition
of the energy between degrees of freedom 
is responsible for the normal thermal conductivity. In fact, we see that, around the
distribution maxima, the particles arrange in the form that hot light particles are surrounded
by cold heavy ones. The energy is then trapped and released in a diffusive way. But we also see
that the release is diffusive when an enough large number of those hot-cold structures
develop. Therefore we think that the mechanism for the thermal resistance 
is somehow cooperative. 

In conclusion, Peierls arguments  have successfully explained the observed thermal
conductivity in solids by applying a perturbative scheme around the lattice harmonic interaction.
The actual belief is that strong anharmonicity is not enough to guarantee a normal
thermal conduction in one dimensional systems. Moreover, it has been proposed that
the key lacking ingredient is that the dynamics of the system 
should  not conserve linear momentum via the existence of local potentials through  the line
(think about particles attached to the one dimensional substrate through some kind of
nonlinear springs). In such a way, local potentials should
act as local energy reservoirs that slow down the energy flow.
These properties, anharmonicity and non-conservation of
momentum, are in some way the ones used on the original Peierls argument.
We have shown a model that does not follow such clean picture.
Although our one dimensional model is nonlinear and it conserves linear momentum 
(with non-zero pressure),
we find that it follows Fourier's law. We think that there are other cooperative mechanisms
that can do the job of the local potentials. Maybe,  systems having  degrees of freedom that
acquires easily energy but releases it in a very long times scale have, in general,
normal thermal conductivity.
In any case, we think that it is worth to explore such possibility.

\vskip 0.5cm
{\it ACKNOWLEDGMENTS-}
 It is a pleasure to acknowledge J.L. Lebowitz, G. Gallavotti, R. Livi, F. Bonetto,
V. Ricci, L. Rey-Bellet, J. Marro and E. Presutti
for useful discussions and criticisms.
This work has been partially supported by 
Junta de Andaluc{\'\i}a, project FQM-165
and by the Ministerio de Educaci\'on:
 projects DGESEIC PB97-0842 and PB97-1080.


\end{document}